\documentclass[preprint]{elsarticle}
\makeatletter
\def\ps@pprintTitle{%
 \let\@oddhead\@empty
 \let\@evenhead\@empty
 \def\@oddfoot{}%
 \let\@evenfoot\@oddfoot}
\makeatother
\usepackage[utf8x]{inputenc}
\usepackage{amsmath,amssymb}

\usepackage{nicefrac}
\usepackage{lineno,hyperref}
\hypersetup{colorlinks = true, allcolors=blue}

\begin{document}

\begin{frontmatter}

\title{Derivative~pricing~as~a~transport~problem: MPDATA~solutions to Black-Scholes-type equations\tnoteref{t1}}
\author[1]{Sylwester Arabas}
\ead{sylwester.arabas@uj.edu.pl}
\author{Ahmad Farhat}

\address[1]{Jagiellonian University, Cracow, Poland}

\tnotetext[t1]{This paper summarises research carrier out at 
  Chatham Financial Corporation Europe}

\begin{abstract}
We discuss in this note applications of the Multidimensional
  Positive Definite Advection Transport Algorithm (MPDATA) to 
  numerical solutions of partial differential equations arising
  from stochastic models in quantitative finance.
In~particular, we develop a framework for solving Black-Scholes-type equations by first transforming
  them into advection-diffusion problems,
  and numerically integrating using an iterative explicit finite-difference approach,
  in which the Fickian term is represented as an additional advective term.
We discuss the correspondence between transport phenomena and financial models,
  uncovering the possibility of expressing the no-arbitrage principle
  as a conservation law.
We depict second-order accuracy in time and space of the embraced numerical scheme. 
This is done in a convergence analysis comparing MPDATA numerical solutions 
  with classic Black-Scholes analytical formul\ae~for the valuation 
  of European options.
We demonstrate in addition a way of applying MPDATA to solve the 
  free boundary problem (leading to a linear complementarity problem) 
  for the valuation of American options. 
We finally comment on the potential the MPDATA framework has with respect to 
  being applied in tandem with more complex models typically used in 
  quantitive finance. 
\end{abstract}

\end{frontmatter}


\section*{Introduction}

MPDATA stands for Multidimensional Positive Definite Advection Transport Algorithm.
The algorithm was introduced in \cite{Smolarkiewicz_1983,Smolarkiewicz_1984} as a robust numerical
  scheme for atmospheric modelling applications.
Thanks to continued research, extensions, and generalisations of MPDATA 
  (see MPDATA review papers \cite{Smolarkiewicz_and_Margolin_1998,Smolarkiewicz_2006}), 
  it has been applied in a wide range 
  of~computational research for numerical integration of partial differential equations 
  describing transport phenomena.
Applications include modelling of
  brain injuries, transport in porous media, sand dune formation, convective cloud systems, 
  operational weather prediction, and studies of climate dynamics and solar magnetohydrodynamics 
  (refer to \cite[sec.~1.2]{Smolarkiewicz_et_al_2016} for a recent review of applications).

The Black-Scholes model \cite{Black_and_Scholes_1973} is a mathematical description of the behaviour of~financial markets
  in which trading occurs in financial assets, as well as derivative financial instruments - contracts whose values
  are dependent on prices of other assets.
The~model gives rise to formul\ae~routinely used in the financial industry to price 
  derivatives.
The~1997 Nobel Prize in economic sciences was awarded to
  contributors to this pricing methodology, Robert Merton and Myron Scholes.

The goal of this paper is twofold.
First, we wish to~attract the mostly-geoscientific MPDATA community
  to applications in~quantitative finance, a domain replete with 
  applications of finite-difference methods (see, e.g., \cite{Duffy_2006}).
Second, we intend to turn the attention of the quantitative finance
  community to a family of~accurate finite-difference solvers 
  possessing characteristics
  that are advantageous in tackling derivative pricing problems:
  conservativeness, high-order accuracy, low numerical diffusion, and
  monotonicity-preserving oscillation-free solutions.
This is in line with the proposal put forward in \cite{Duffy_2004}
  to investigate robust and effective numerical schemes documented in the
  computational fluid dynamics literature as alternatives to commonly used numerical
  schemes in financial engineering, with the aim of ``improving the
  finite difference methods gene pool as it were.''
To these ends, leveraging the mathematical equivalence between
  Black-Scholes-type models and transport models,
  we detail applications of MPDATA
  to numerically reproduce the analytical solution of a celebrated 
  benchmark problem --- the~Black-Scholes formula for pricing of European options ---
  and to numerically solve the associated free boundary problem
  arising in the valuation of American options.

With the aim of catering to both communities, we begin this note with a brief
  introduction to both the Black-Scholes model and the MPDATA solver.
We purposefully
  include explanations of terms that can be considered elementary in their respective domains.
Also included is a discussion of the variable transformation
  that converts the Black-Scholes equation into a homogeneous
  advection-diffusion equation.
The background section is followed by a description of a sample application 
  of MPDATA for pricing a financial instrument composed of
  European options.
First, we detail the numerical solution procedure, and discuss 
  the results for a single simulation.
Second, we corroborate results of multiple simulations with analytical solutions 
  in an analysis quantifying the rate of convergence of the numerical solutions for both
  upwind and MPDATA schemes.
We conclude this note by highlighting the potential MPDATA has for further 
  applications in finance.
In the appendix, we present an extension of the developed framework
  for pricing American options.


\section*{Background} 

\subsection*{The Black-Scholes model in a nutshell}

A common ansatz in financial market modelling is that the price $S$ of an asset follows a continuous-time 
  lognormal diffusion process known as geometric Brownian motion\footnote{The ansatz is credited to, among others, Osborne \cite{Osborne_1959} who referred to it 
  as the hypothesis that price and value are related by the Weber-Fechner law.}.
This process is modelled by the stochastic 
  differential equation (SDE):
\begin{equation}\label{eq:gbm}
  dS = S(\mu dt + \sigma dw) 
\end{equation}
where $\mu$ and $\sigma$ are constants denoting the expected instantaneous return on investment in the asset and 
  the asset price volatility, respectively, $t$ denotes time and $w$ is a Wiener process (also called a Brownian motion).
This simple model embodies the fact that what matters to investors is the rate of return on 
  their investment in an asset, and not the change in the asset price (in which case, the $S$ term 
  would be dropped from the right-hand side of eq.~\ref{eq:gbm} as in the Bachelier model 
  \cite{Bachelier_1900}, a seminal fin-de-siècle starting point for mathematical finance and, 
  it is noteworthy, a seminal work in the theory of Brownian motion itself; for discussion see \cite{Schachermayer_and_Teichmann_2007}).
Furthermore, this model entails two propositions:
  (i) in the limit where the volatility is 
  negligible, the investment in the asset mimics a deposit with interest rate $\mu$;
  (ii) in the opposite limit where $\mu$ is negligible, the return on the investment is 
  random with a normal distribution.  
  
The Black-Scholes model assumes that the modelled asset price follows a geometric Brownian motion.
Key among the other model assumptions are that the rate of return on a riskless investment
  is fixed and given by the so-called ``risk-free interest rate'' 
  (a high-rated government bond can be thought of as a 
  surrogate for the idealised riskless investment),
  and that there are no arbitrage opportunities (precluding the 
  possibility of riskless returns in excess of the 
  the~risk-free interest rate).

Suppose, in the Black-Scholes model, that a derivative instrument is also traded
  in the market.
For instance, a ``European call option'' on an asset (e.g., a stock) is~a~type of~a~derivative that 
  gives its holder the right, but not an obligation, to purchase the underlying 
  asset on a specified future date at a specified price.
An American option differs from a European option only in that it can be exercised at any time
  prior to its date of expiration (cf. Appendix).
Given an asset in the Black-Scholes model whose price process is given by $S$, 
  the~aim is~to~discover the price of a derivative contingent on $S$.  

Let $f(S,t)$ be the value of an option dependent on the asset price~$S$ at~time~$t$.
Since $S$ follows a Wiener process, the change in $f$ can be expressed using It\^o's lemma as:
\begin{equation}\label{eq:Ito}
    df 
  	   = \left(\frac{\partial f}{\partial t} + \mu S\frac{\partial f}{\partial S} + \frac{1}{2}\sigma^2 S^2\frac{\partial^2 f}{\partial S^2}\right)dt + \sigma S\frac{\partial f}{\partial S} dw
\end{equation}

The crucial observation is that the asset price $S$ and the option value $f$ have the~same 
  source of randomness, associated with the Wiener process $w$.  
Thus, one can construct a suitably weighted portfolio 
  by selling one unit of the option and holding as much, 
  $\Delta_t$, of the underlying asset
  so as to eliminate the randomness and make the portfolio riskless.  
In finance, risk reduction is referred to as hedging.
The portfolio value $\Pi(S, t)$ is given by 
\begin{equation}\label{eq:pi}
  \Pi = -f + \Delta_t S
\end{equation}
Substituting from eq.~(\ref{eq:gbm}) and eq.~(\ref{eq:Ito}), we have
\begin{equation*}
    -df + \Delta_t dS 
    	= \left[-\frac{\partial f}{\partial t} 
		- \frac{1}{2}\sigma^2 S^2\frac{\partial^2 f}{\partial S^2} 
		+ \left(\Delta_t - \frac{\partial f}{\partial S}\right)\mu S\right]dt 
		+ \left(\Delta_t - \frac{\partial f}{\partial S}\right)\sigma S dw
\end{equation*}
showing that the only stochastic contribution to the portfolio value at time $t$ is given by
$\int_0^t \left(\Delta_u - \frac{\partial f}{\partial S_u}\right)\sigma S_u dw_u$.
Thus, by adopting the so-called delta-hedging strategy, with the proportion $\Delta_t$ 
  of the asset held at time $t$ assumed to be locally constant and equal 
  to~$\left.\frac{\partial f}{\partial S}\right|_t$, the~portfolio is~instantaneously riskless. 
A riskless portfolio must evolve according to the risk-free interest rate $r$: 
\begin{equation}\label{eq:noarb}
  d\Pi = \left(-\frac{\partial f}{\partial t} - \frac{1}{2}\sigma^2 S^2\frac{\partial^2 f}{\partial S^2} \right)dt = \Pi r dt
\end{equation}
Substituting from eq.~(\ref{eq:pi}) into eq.~(\ref{eq:noarb}) yields the celebrated 
  Black-Scholes equation \cite{Black_and_Scholes_1973}:
\begin{equation}\label{eq:BS}
  \frac{\partial f}{\partial t} + rS \frac{\partial f}{\partial S} + \frac{\sigma^2}{2} S^2 \frac{\partial^2 f}{\partial S^2} - rf = 0
\end{equation}
The derivation of eq.~(\ref{eq:BS}) hinged on the elimination of the 
  stochastic term, reducing the SDE to a partial differential equation (PDE).
It is worth noting that there exists an alternative approach of casting the 
  option valuation problem in PDE form via the Martingale pricing theorem, 
  using the Kolmogorov forward equation (or Fokker-Planck equation) 
  \cite{Paul_and_Baschnagel_2013}.

\subsection*{Derivative pricing as a transport problem}

The Black-Scholes equation can be transformed into a homogeneous advection-diffusion (convection-diffusion, scalar transport) equation
  using the following variable substitution:
\begin{equation}
\label{eq:changeVariable}
\left\{
\begin{array}{llr}
	\psi&= e^{-rt}f(S,t) \\
    x&=\ln S \\
    u&= r-\frac{\sigma^2}{2} \\
    \nu&= -\frac{\sigma^2}{2}
\end{array}    
\right.
\end{equation}
leading to:
\begin{equation}\label{eq:adv}
  \frac{\partial \psi}{\partial t} + u \frac{\partial \psi}{\partial x} - \nu\frac{\partial^2 \psi}{\partial x^2} = 0
\end{equation}
The Black-Scholes methodology relies on solving a~terminal value problem (hence the negative sign of $\nu$); 
  the substitution (\ref{eq:changeVariable}) can be extended to~lead to~an~initial value problem by 
  introducing $\tau = T - t$ as in \cite[eq.~5.68-5.71]{Joshi_2008}.
The connection between the Black-Scholes equation and convection-diffusion 
  equations has been mentioned in the literature; see e.g. \cite[sec.~1.1]{Morton_1996} in which 
  the notion of ``financial drift'' is used.

Adapting a general technique expounded in \cite[Sec.~3.2]{Smolarkiewicz_1986} and \cite{Sousa_2009,Smolarkiewicz_and_Szmelter_2005}
  (as well as in \cite{Cristiani_2015} for the case of diffusion-only problem),
  eq.~(\ref{eq:adv})
  can be rearranged to mimic an advection-only problem,
  assuming $u$ and $\nu$ constant in $x$:
\begin{equation}\label{eq:advonly}
  \frac{\partial \psi}{\partial t} + \frac{\partial}{\partial x}\left[\left(u - \frac{\nu}{\psi}\frac{\partial \psi}{\partial x} \right) \psi \right] = 0
\end{equation}
which we will leverage in numerical solutions of eq.~(\ref{eq:adv}).
Equations (\ref{eq:noarb}) and (\ref{eq:noarb}) followed from the 
  no-arbitrage condition, which embodies the assumption that a 
  riskless portfolio cannot have returns in excess of the risk-free interest rate. 
In eq.~(\ref{eq:advonly}), this no-arbitrage condition is cast in the form of 
  a conservation law.  

Noteworthy, the prevalent approach to solving the Black-Scholes equation is
  to transform it to a heat equation, e.g. by amending the variable substitution~(\ref{eq:changeVariable})
  by setting $x=\ln S + (r - \sigma^2/2)\tau$ \cite[eq.~5.72]{Joshi_2008}.
Such approach, which can be viewed as a transformation of the transport problem from an Eulerian into a Lagrangian 
  frame of reference, leads to the elimination of the advective term and
  enables the derivation of an analytical solution of the Black-Scholes equation.
In contrast, the approach embodied in eq.~(\ref{eq:advonly}), which leads to the elimination of the Fickian term,
  facilitates numerical integration by not introducing time-dependent coordinate transformations
  and by allowing for consistent discretisation of both the advective and Fickian fluxes.

Commenting on the variable substitution~(\ref{eq:changeVariable}), we note that
 $x =\ln(S)$ transforms the~Black-Scholes equation into a constant-coefficient advection-diffusion equation with a~source term.
Introducing $\psi(x,t)$, in financial terms the present (discounted) value of~the~option, reduces the equation
  to a homogeneous one.
This is akin to the incorporation of adiabatic cooling/heating in atmospheric heat budget equations, not through 
  the use of a source term, but rather
  through the introduction of potential temperature.
One may note a curious analogy in the descriptive definitions of the two quantities.
Potential temperature, linked with the entropy of an ideal gas (see \cite{Bauer_1908} for a historical perspective on its introduction),
  is commonly described as the temperature a parcel of air would have if brought adiabatically to a base level ``zero''. 
The discounted option price~$\psi$ represents the value that the option would have if brought from 
  its state at a future time~$t$ to~the~present time~$t=0$.

Equation (\ref{eq:adv}) (and its generalisations) is a staple in geoscientific research, 
  where it~is~used for~modelling transport
  phenomena.
For instance, it can depict the~transport in the atmosphere
  of a pollutant concentration field $\psi$ by wind of velocity $u$ subject to diffusion 
  with coefficient $\nu$.
In finance, the key application of eq.~(\ref{eq:adv}) is~to~solve, backwards-in-time, 
  for the current price of the option $f(S_0, 0)=\psi(\ln(S_0), 0)$,
  where $S_0$ is the current price of the underlying asset.
The terminal condition (starting point for the solver) is~given by the so-called payoff function 
  $f(S, T)=\psi(\ln(S),T)$, defining the type of
  derivative contract under consideration (option to buy the asset, option to sell the asset, combination of such options, etc.).

Making a heuristic physical analogy, we note that eq.~(\ref{eq:adv}) in our current financial context
  governs the transport of the option price $f$ discounted to its present value $\psi$. 
The second term of eq.~(\ref{eq:adv}), which governs the advection of the quantity of interest, 
  $\psi$, incorporates a velocity $u$ at which $\psi$ is moving. 
Noting that since the underlying process $S$ is~governed by a geometric Brownian motion, and that the spatial 
  variable in eq.~(\ref{eq:adv}) is~$x=\ln(S)$, the solution of the geometric Brownian motion 
  SDE~(\ref{eq:gbm}) with $\mu=r$ implies that the drift on $x$ is precisely 
  $(r-\frac{\sigma^2}{2})$ (here the replacement of $\mu$ by $r$ is justified on~the~grounds 
  of risk-neutral pricing for which we omit the details). 
This~explains the form of the advective term in eq.~(\ref{eq:adv}).
Moreover, It\^o's lemma implies that any twice-differentiable scalar function of $S$ and $t$ 
  will have a diffusivity coefficient precisely $\frac{\sigma^2}{2}$,
  which explains the Fickian term in eq.~(\ref{eq:adv}).
Thus, eq.~(\ref{eq:adv}) could be viewed as~describing the transport
  of the discounted option price $\psi$ over the space $x=\ln(S)$, 
  with the dynamics of $S$ conferring an advective velocity given by $u$ 
  and a diffusivity coefficient given by $\nu$. 
Analysis of this type serves to elucidate derivative pricing dynamics when more sophisticated 
  SDEs govern the behavior of the underlying assets. 

\subsection*{MPDATA in a nutshell}

MPDATA is a family of numerical schemes for solving transport problems.
Its basic formulation numerically integrates equations of the form:
\begin{equation}\label{eq:advtilde}
  \frac{\partial \psi}{\partial t} + \frac{\partial}{\partial x}\left(v \psi \right) = 0
\end{equation}
  after discretisation in~time 
  $t\in\Delta t \cdot \{0, \ldots, n, n+1, \ldots\}$ and space $x\in\Delta x \cdot \{0, \ldots, i, i+1, \ldots\}$,
  where $\Delta t$ is~the~timestep and $\Delta x$ is~the~gridstep.
It is an iterative, explicit-in-time finite-difference algorithm in which 
every iteration takes the form:
\begin{equation}\label{eq:upwind_a}
  \psi_i^{n+1} = \psi_i^n - \left[
    F(\psi_i^n, \psi_{i+1}^n, \mathcal{C}_{i+\nicefrac{1}{2}}) - 
    F(\psi^n_{i-1}, \psi_i^n, \mathcal{C}_{i-\nicefrac{1}{2}})
  \right]
\end{equation}
where the two instances of the function $F$ depict the fluxes of the transported quantity from grid cell 
  $i$ to $i+1$ and from grid cell $i-1$ to $i$, respectively; $\mathcal{C}$ is the Courant number defined
  as $v\frac{\Delta t}{\Delta x}$; 
  fractional indices (i.e., $i\pm\nicefrac{1}{2}$) indicate that $\mathcal{C}$ is evaluated at grid cell boundaries,
  whereas integer indices indicate evaluation at cell centers (see Fig.~3 in~\cite{Jaruga_et_al_2015}).
$F$~is defined as:
\begin{equation}\label{eq:upwind_b}
  F(\psi_L, \psi_R, \mathcal{C}) = \max(\mathcal{C},0) \cdot \psi_L + \min(\mathcal{C},0) \cdot \psi_R
\end{equation}
Introducing $v$ and $\mathcal{C}$ to denote the advective velocity and Courant number, 
  respectively, serves to distinguish their general meaning (including the possibility of time and space dependence)
  from the particular context introduced in the preceding section (where $u$ and $C=u\frac{\Delta t}{\Delta x}$ are constant).

The first iteration of MPDATA is equivalent to the so-called upwind (donor-cell, upstream)
  integration method, which suffers from extensive ``numerical diffusion'', i.e., smoothing of the signal
  (for some poingnant remarks on the issue and a vivid recount of the relevant controvercies, see \cite{Leonard_1979}).
The term ``numerical diffusion'' stems from the fact that when the numerical approximation 
  of eq.~(\ref{eq:advtilde}),
  expressed by eq.~(\ref{eq:upwind_a}-\ref{eq:upwind_b}), is analysed through the modified equation approach,
  the leading terms of the truncation error estimate can be expressed in the form of $K\partial_x^2 \psi$.
The nub of MPDATA lies in expressing this truncation error estimate 
  as an additional advective term (vide eq.~\ref{eq:advonly}).
This additional numerical-diffusion-reversing term is integrated in a subsequent iteration,
  using the very same conservative and positive-definite upwind scheme.
As a result, the truncation error estimate is subtracted from the solution.
Solving $\partial_x (u' \psi) = K \partial^2_x \psi$ for $u'$ and discretising, for the most
  basic formulation of MPDATA, gives
  the following so-called antidiffusive Courant number to be used in the corrective iteration of MPDATA:
\begin{equation}\label{eq:antidif}
  \mathcal{C}'_{i+\nicefrac{1}{2}} = (|\mathcal{C}_{i+\nicefrac{1}{2}}| - \mathcal{C}_{i+\nicefrac{1}{2}}^2) A_{i+\nicefrac{1}{2}}
\end{equation}
where
\begin{equation}\label{eq:A}
  A_{i+\nicefrac{1}{2}} = \frac{\psi_{i+1} - \psi_{i}}{\psi_{i+1} + \psi_{i}}
\end{equation}
with the values of $\psi$ corresponding to results from the first iteration. 
Even in the basic formulation of MPDATA, the corrective iteration makes the scheme second-order accurate in~time and space.
Subsequent iterations reduce the magnitude of the error while maintaining second-order accuracy.
Extension of the analysis, by taking into account higher-order terms in the 
  Taylor expansion, leads to construction of higher-order MPDATA schemes
  (see \cite{Waruszewski_et_al_2018} for a fully third-order variant).

MPDATA is by design sign-preserving (i.e., a non-negative initial state leads to
  a non-negative solution), which is a non-trivial property among higher-order advection schemes.
This is an essential prerequisite in such applications as 
  option pricing in finance or pollutant advection in geoscience;
  the quantities in question need to remain non-negative for the solution
  to make sense: negative pollutant concentrations are unphysical, and the fact that option holders are not 
  obliged to exercise implies that option values are non-negative.
There are several extensions of MPDATA of particular applicability in quantitative finance, including the
  non-oscillatory option that ensures the elimination of spurious oscillations in the solution
  using a technique derived from the flux-corrected-transport methodology discussed in the context
  of solutions to derivative pricing problems in \cite{Zvan_et_al_1998}.
While in the above outline of the derivation of MPDATA a one-dimensional 
  problem was taken into consideration, let us point out for clarity that
  in~the~vast majority of its applications, MPDATA had been employed
  for solving multi-dimensional problems.
The multi-dimensional formulations of advective velocities feature cross-dimensional terms,
  which distinguishes MPDATA from dimensionally-split schemes.

A~significant subset of the MPDATA family of algorithms 
  has recently been implemented in C++ and released as an open-source reusable 
  library called libmpdata++ \cite{Jaruga_et_al_2015}.
The~example simulations presented in the following section were implemented using libmpdata++.

\section*{European option valuation using MPDATA}

\subsection*{Benchmark problem}

The need for finite-difference methods clearly arises when no analytical solutions 
  are available. 
Nevertheless, in order to demonstrate the MPDATA numerical framework, 
  we price a so-called corridor under the Black-Scholes model.
This allows us to corroborate the numerical results 
  against the Black-Scholes analytical pricing formul\ae.
The priced corridor is a compound instrument composed of two European options: 
  a bought (long) option to sell an underlying asset at price $K_2$
  and a sold (short) option to sell the asset at price $K_1$, where
  $K_1$ and $K_2$, referred to as the strike values, satisfy $K_1<K_2$%
\footnote{
The corridor here is a financial instrument designed to reduce (hedge against) 
  a decrease in the underlying asset price below $K_2$ 
  through the bought option while offsetting the cost of the bought option by
  the simultaneous sale of the lower-strike option.
More specifically, if the value of the underlying asset price $S$ at the 
  time of the option expiry is above $K_2$, the corridor payoff is zero 
  (neither of the options will be exercised) -- this is the range of values of $S$
  for which the corridor owner does not require any protection.
If the value of $S$ is between $K_1$ and $K_2$, the corridor payoff is proportional to
  the difference $(K_2-S)$ -- in this range the corridor effectively eliminates the consequences
  of underlying price movements.
For any value of $S$ less than $K_1$, the corridor payoff stays constant at 
  $(K_2-K_1)$,
  thereby providing no protection against price decreases below $K_1$.
An example rationale for such hedging strategy is when 
  little probability is ascribed to the event of the underlying price 
  falling below $K_1$.
}.
The payoff function for such corridor is:
\begin{equation}\label{eq:payoff}
  f(S, T) = \max(K_2-S, 0) - \max(K_1-S, 0)
\end{equation}
The payoff function has a vanishing first derivative when $S<K_1$ or $S>K_2$,
  which makes it easier to apply standard open boundary conditions at the edges
  of the computational domain.
This is why the corridor example is an apt
  elementary case from the perspective of the finite-difference solver.

\subsection*{Numerical solution procedure}

Pricing the corridor using MPDATA is done as follows:
\begin{itemize}
  \item{The terminal condition defined by $\psi(\ln(S), T)$ is evaluated by discretising the~payoff function discounted by the factor $e^{-rT}$.}
  \item{The numerical integration of the transport equation is carried out by solving from $t=T$ to $t=0$ (i.e., with negative timesteps of magnitude $\Delta t$).}
  \item{The value of $\psi(\ln(S_0), 0)$ is the sought after price of the corridor, where $S_0$ is~the~present price of the underlying asset.
    Note that for $t=0$, the exponential factor in $\psi$ is equal to 1.}
\end{itemize}
Unmodified code of libmpdata++ is used for handling the integration of the advective term in 
  the transport equation.
Custom code is used to discretise the~Fickian term
  as an addition to the advective velocity.

The computational grid is chosen by dividing $T$ into $n_t$ equally-sized timesteps $\Delta t$, 
  and by laying out $n_x$ grid points using equally-sized gridsteps $\Delta x$.
The values of $\Delta t$ and $\Delta x$ control the accuracy of the solution, and the ranges of their values are
  bound by the stability constraint of MPDATA:
\begin{equation}
\mathcal{C} = \left|r-\frac{\sigma^2}{2}+\frac{\sigma^2}{\Delta x}A\right|\frac{\Delta t}{\Delta x} < \frac{1}{2}
\end{equation}
where the A term defined by (\ref{eq:A}) stems from the discretisation of the Fickian term,
  and the limit of $1/2$ applies in the case of divergent velocity field (non-constant $v$ in the one-dimensional case);
  see discussion in \cite{Smolarkiewicz_1984,Smolarkiewicz_1986}.
Noting the boundedness of A for positive-definite $\psi$,
  the stability condition can be approximated with:
\begin{equation}\label{eq:bigoh}
  \lambda^2 = \frac{1}{\sigma^2} \frac{\Delta x^2}{\Delta t} \gtrsim 2
\end{equation}
where $\lambda^2$ was introduced in accordance with notation from \cite{Kwok_and_Lau_2001}
($\lambda^2$ is inversely proportional to the mesh ratio $R$ discussed in
  \cite[Sec.~II.B]{Geske_and_Shastri_1985}, to~the~parameter~$w$
  discussed in \cite[Sec.~III.B]{Hull_and_White_1990}, to the
  diffusion number $r$ defined in \cite{Hogarth_et_al_1990}, 
  and~to~the mesh Fourier number $\mu$ defined in \cite{Sousa_2009}).
As discussed in \cite[sec.~3.2]{Smolarkiewicz_1986}, the constraint~(\ref{eq:bigoh}) is
  twice more stringent than for the standard first-order-in-time FTCS scheme.

Notably, employing consistent discretisation for both the advective and Fickian terms
  in eq.~(\ref{eq:adv}) ensures that, for linear payoffs (as in the case of forward contracts), the terms featuring $\sigma$
  cancel out in the numerical solution.
This is not the case if, for instance, an upwind scheme is used for representing
  the advective term while a central-difference is employed in the discretisation of the Fickian term.
  
\subsection*{Analytical solution}

In their seminal paper \cite{Black_and_Scholes_1973}, Black and Scholes gave
  solutions to eq.~(\ref{eq:BS}) for payoff functions associated with European options.
Following their results, the value (at $t=0$) of the corridor is given by:
\begin{equation}\label{eq:psi_a}
  f(S_0, 0) = p(S_0, K_2) - p(S_0, K_1)
\end{equation}
where $p(S_0, K)$ is the Black-Scholes formula for the price of a ``put'' option: 
\begin{equation}\label{eq:bs_put}
  p(S_0, K) = -S_0 N(-d_1) + K e^{-rT} N(-d_2(K))
\end{equation}
where $d_1(K)=\left[ \ln(S_0/K) + (r+\sigma^2/2)T\right]/(\sigma \sqrt{T})$,
  $d_2(K)=d_1(K)-\sigma \sqrt{T}$ and $N(x)$ denotes the standard normal cumulative distribution function.

Interestingly, taking $K\!=\!1$, $\sigma^2\!=\!2$ and $r\!=\!0$, we have an
  equivalence with the ``standard model for the transport of an~unreactive solute in a soil column''
  used in a finite-difference scheme analysis~in~\cite{Hogarth_et_al_1990}.

\subsection*{Results}

\begin{figure}[!h]
  \caption{{\bf Corridor valuation.}
  Comparison of a numerical solution obtained with MPDATA with the corresponding analytical solution (i.e.,~the~Black-Scholes formula).
  Instrument parameters: a sold option with strike $K_1=75$ and a
  bought option with strike $K_2=125$,
  6-month tenure (time~to~expiry), risk-free rate $r=0.8\%$, volatility $\sigma=0.6$.}
  \begin{center}
    \includegraphics[width=\textwidth]{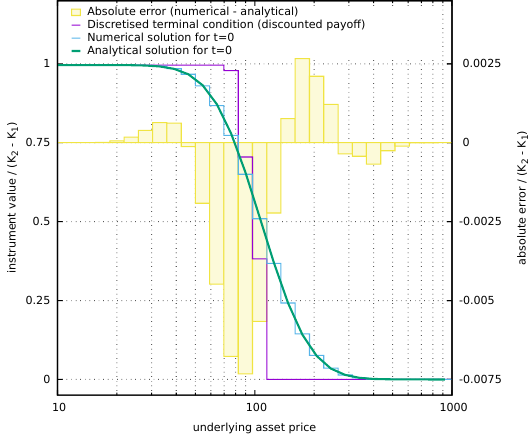}
  \end{center}
  \label{fig1}
\end{figure}

In Fig.~\ref{fig1}, an example numerical solution (in blue) is presented alongside the discretised terminal condition (in purple), 
  the analytical solution (in green) and the difference between the~two (in~yellow).
Parameters of the corridor are given in the figure caption.
The abscissa corresponds to the value of~the~underlying asset $S$; since in the case of the corridor 
  it~is~an interest rate, it is expressed in percents.
The left ordinate denotes the value of~the~derivative $f$, expressed
  as a percentage of the notional $N$.
The right ordinate denotes the absolute error, expressed in percentage points.
The solver states at $t=T$ (terminal condition) and at $t=0$ are plotted with histogram-like curves to
  depict the computational grid layout.
The solution was obtained with $\lambda^2=2$ and $C=u\frac{\Delta t}{\Delta x}\approx0.04$, resulting in ca. 10 timesteps and ca. 30 grid elements.

The numerical solution was obtained with the following settings of libmpdata++
  (consult \cite{Jaruga_et_al_2015} for details):
  one corrective iteration,
  non-oscillatory, infinite-gauge, and divergent-flow options enabled,
  a choice for which the highest convergence rate in time was observed. 
Figure~\ref{fig1} qualitatively depicts the match between the numerical and analytical solutions.
It shows that the error is smallest near the domain boundaries, confirming that the 
  domain extent is sufficient for the given parameters.
The solution does not feature values below zero (the minimum of the initial condition), which 
  illustrates the positive definiteness of MPDATA.
The solution does not feature values above the maximum of the initial condition which in turn 
  demonstrates the conservativeness and monotonicity (non-oscillatory character) of the scheme.

\begin{figure}[!h]
  \caption{{\bf Solution accuracy in terms of the spatial discretisation.}
    Truncation error as a function of the Courant number $C=u\frac{\Delta t}{\Delta x}$ which, for fixed $\lambda^2$, is 
      proportional to~the~gridstep.
    Thin lines correspond to the basic upwind scheme (first iteration of~MPDATA only), 
    thick lines correspond to results obtained with one corrective iteration of MPDATA.
    Three datasets plotted for three different values of $\lambda^2$.
    The dotted and solid black lines depict the slopes corresponding to first-order and second-order convergence.
  }
  \begin{center}
    \includegraphics[width=.72\textwidth]{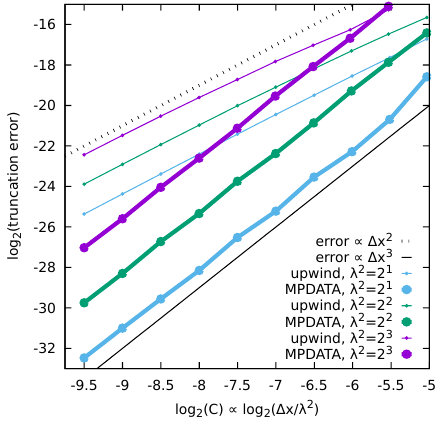}
  \end{center}
  \label{fig2}
\end{figure}
\begin{figure}[!h]
  \caption{{\bf Solution accuracy in terms of the temporal discretisation.}
    Truncation error as a function of the $\lambda^2$ parameter which, for fixed $C$, is 
      proportional to~the~timestep.
    Three datasets plotted for three different values of $C$ (values given approximately as the 
      solution procedure adjusts the requested value so~that~the~number of timesteps is an integer).
    Other plot elements are as in Fig.~\ref{fig2}.
  }
  \begin{center}
    \includegraphics[width=.72\textwidth]{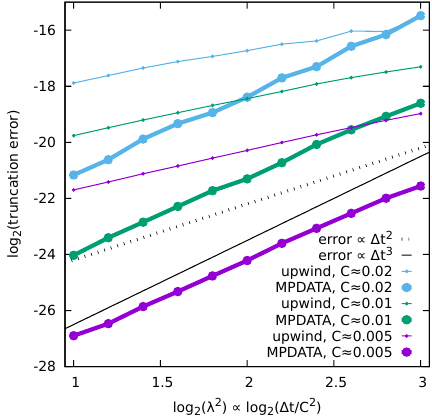}
  \end{center}
  \label{fig3}
\end{figure}

A quantitative analysis of the errors arising from the numerical integration is summarised in Figs~\ref{fig2}-\ref{fig3}.
The accuracy of the solution is quantified using an $L_2$ measure of the average error per timestep and per gridstep,
  defined following \cite{Smolarkiewicz_1984} as:
\begin{equation}\label{eq:error}
  E = \sqrt{\left.\sum\limits_{i=1}^{n_x}\left[ \psi_n(x_i) - \psi_a(x_i) \right]^2 / (n_x \cdot n_t) \right|_{t=0}}
\end{equation}
where $\psi_n$ is the numerical solution, and $\psi_a$ is the analytical one given by eq.~(\ref{eq:psi_a}). 
In Fig.~\ref{fig2}, the base-2 logarithm of $E$ is plotted against the base-2 logarithm of $C=u\frac{\Delta t}{\Delta x}$ for~several settings of $\lambda^2$.
Thick lines represent solutions obtained with two iterations (labelled as MPDATA), thin lines represent solutions 
  obtained with a single-pass scheme,
  i.e., the basic upwind algorithm.
All other solution parameters were set as~in~the~example depicted in Fig.~\ref{fig1}.

Since, for a given value of $\lambda^2$, $C$ is proportional to the gridstep $\Delta x$, the slopes
  of the plotted curves depict how the results converge when refining the spatial discretisation.
To facilitate interpretation, two additional curves were plotted, depicting the theoretical slopes
  for first-order and second-order accuracy.
Figure~\ref{fig2} confirms that for the problem at hand, and for the three presented settings of $\lambda^2$, 
  MPDATA is of second-order accuracy in space,
  improving over the close to first-order accurate solutions obtained with the upwind scheme.

The rate of convergence of the numerical solution to the analytical one as a function of the timestep
  is depicted in Fig.~\ref{fig3}, constructed similarly to Fig.~\ref{fig2}, with base-2 logarithm
  of $\lambda^2$ on the abcissa.
Since for a given value of $C$, $\lambda^2$ is proportional to~the~timestep~$\Delta t$, the plotted curves
  depict the order of accuracy in time.
In all presented cases, the MPDATA solutions are of second-order accuracy in time, while the 
  convergence rate of the upwind solutions in time is below first order.
  
\section*{Summary and prospects}

This work was intended to serve as a springboard for applications of MPDATA 
  in quantitative finance.
In this domain, the MPDATA family of numerical schemes appears to be particularly promising and 
  adaptable for solving PDEs arising in derivative pricing problems. 
It possesses particularly appealing properties in terms of:
\begin{itemize} 
  \item positive definiteness (non-negativity of option price solutions by design),
  \item monotonicity (no spurious oscillations in the solutions),
  \item conservativeness and high-order accuracy (second-order in time and space for the basic MPDATA), 
  \item multidimensionality (superior to dimensionally-split schemes; applicable to problems giving rise to multi-dimensional PDEs), 
  \item robustness (explicit and hence trouble-free to implement, and apt to parallelisation via domain decomposition).
\end{itemize}

The prospects for the use of the MPDATA framework in quantitative finance lie in its applications
  in more sophisticated contexts.
We describe in the Appendix, for instance, an extension of the developed framework to handle
  valuation of American options, for which finite difference methods are applied in the industry.
Other potential applications include 
  problems modelled with multi-dimensional PDEs such as 
  in stochastic volatility models and 
  in pricing derivatives incorporating dependence on the history of underlying processes (i.e., path-dependent derivatives, e.g., so-called Asian options), 
  and in pricing of other multi-factor (e.g., multi-asset) derivatives.

\appendix

\section{American option valuation using MPDATA}

\subsection*{Problem formulation}

American options differ from the European ones only by allowing the holder
  to exercise at any time prior to expiry.
Pricing of American options leads to a free boundary problem since there is a boundary, not known in advance,
  that for any time $t$ separates the regions where it is either optimal to continue holding
  the option or optimal to exercise it immediately \citep[see e.g.][Sec.~12.3]{Joshi_2008}.
Under the herein embraced assumptions, notably without considering the so-called costs of carry \citep[][sec.~5.10]{Joshi_2008},
  the boundary is unique (for a discussion of the more general setting, in
  which the boundary can bifurcate, see \citep{Battauz_et_al_2015}).
In deriving eq.~(\ref{eq:noarb}), it was mentioned that the riskless portfolio $\Pi$ must evolve
  according to the risk-free interest rate $r$.
This is because, over the lifetime of the option, if the portfolio had a rate of return
  less than $r$, an investor holding the portfolio would sell it for $\Pi$,
  invest $\Pi$ at the rate $r$, and buy the portfolio back at a later stage, making riskless profit.
If the rate of return is greater than $r$, an investor would borrow $\Pi$ to buy the portfolio,
  and then sell it at a later time, paying back the debt and making riskless profit.
The difference in the case of American options is that the latter case does not hold:
  in the intervening time between buying the portfolio and selling it, the investor runs the
  risk that the sold option would be exercised against them at any moment, changing the
  rate of return of the portfolio.
This implies that the price process $f$ for American options only satisfies
\begin{equation}\label{eq:AmerBS}
   -\Big(\frac{\partial f}{\partial t} + rS \frac{\partial f}{\partial S} + \frac{\sigma^2}{2} S^2 \frac{\partial^2 f}{\partial S^2} - rf\Big) \geq 0
\end{equation}
On the other hand, at any time $t$, $f(S,t) \geq P(S,t)$, where $P(S,t)$ is the payoff function of 
  the option if it is exercised at time $t$. 
This is because of the no-arbitrage condition: if $f(S,t) < P(S,t)$, the holder of the option would 
  exercise immediately, collect the payoff, and then use it to buy the option back at $f(S,t)$. 
Thus, for a standard put option with strike $K$,
\begin{equation}\label{eq:AmerConstraint}
   f(S,t) - (K-S) \geq 0
\end{equation}
When $f(S,t) = P(S,t)$, the option holder must exercise immediately - continuing to hold the option
  puts the holder at risk of loss.
Replacement in eq.~(\ref{eq:AmerBS}) shows that strict inequality holds in such case.
On the other hand, when $f(S,t) > P(S,t)$, the holder of the option should continue to hold it, 
  and in this case, equality holds in eq.~(\ref{eq:AmerBS}). 
This, together with the terminal condition
\begin{equation}\label{eq:term_cond_Amer}
   f(S,T) = \max(K-S,0)
\end{equation}
  describes a so-called 
  linear complementarity problem, where strict inequality holds in at most one 
  of eq.s~(\ref{eq:AmerBS}) and (\ref{eq:AmerConstraint}).

\subsection*{Numerical solution procedure}

A simple (non-second-order) yet robust way to represent the linear complementarity problem
  in the numerical framework embraced herein is to supplement
  the solved transport equation with a source term representing the linear complementarity problem,
  and to integrate according to the following recipe: 
\begin{equation}
  \psi^{*} = \text{MPDATA}(\psi^n)
\end{equation}
\begin{equation}
  R^n=\frac{\max\left(\psi^*, f(S,T)\exp(-rt^{n+1})\right) - \psi^*}{\Delta t}
\end{equation}
\begin{equation}
  \psi^{n+1} = \psi^* + \Delta t R^n
\end{equation}
The source term $R$ effectively represents a limiter on the time derivative of $\psi$ (and hence, the time derivative of $f$).
For a recent discussion of the formulation of the free boundary problem in terms of 
  a linear complementarity constraint on the time derivative of $f$, as opposed to $f$ itself, 
  see \citep{Bokanowski_and_Debrabant_2018}.

To note, an upwind-based numerical solution for the American option valuation problem
  was previously studied in \cite{Vasquez_1998}. 
That study, however, did not involve the transformation of the Black-Scholes equation into a
  constant-coefficient advection-diffusion equation.

\subsection*{Results}

We corroborate the numerical integration
  results for American put option prices against the estimates obtained from the approximate analytical formula
  derived in \cite{Bjerksund_and_Stensland_1993}.
The payoff function for the put option is given by eq.~(\ref{eq:term_cond_Amer}).
A log-linear extrapolation for both $\psi$ and its derivative is used for the 
  boundary condition in this case.
The grid is chosen to include the point corresponding to the spot price $x=\ln(S_0)$.

Results of a series of integrations carried out for different option tenures $T$ and
  spot prices $S_0$, and with three settings for the Courant number $C=u\frac{\Delta x}{\Delta t}$, are given in Table~\ref{tab}.
All simulations were carried out with $\lambda^2=2$, $\sigma=0.2$, $r=0.08$ and $K=100$, on
  a grid covering the range of $S\in(0.05, 500)$.
The obtained option valuations converge with decreasing $C\sim\Delta x/\lambda^2$  
  to the prices obtained with the Bjerksund and Stensland formula (column labelled BS93).
The error measures computed following eq.~(\ref{eq:error}), and given in the columns labelled $\log_2(E)$, 
  depict a convergence rate in $\Delta x$ not less than first order.
Results obtained with the upwind scheme (not shown) are characterised by larger
  errors than with MPDATA, but comparable convergence rate, illustrating that the
  convergence rate is constrained by the treatment of the linear-complementarity condition.
For reference, prices of European put options with the same parameters, obtained with
  the analytical Black-Scholes formula (\ref{eq:bs_put}), are given in the last column.

\begin{table}
  \caption{\label{tab}
     Prices of American options obtained numerically compared with the Bjerksund and Stensland analytical formula (BS93), and
       with the prices of corresponding European options. Notation: $T$~--~option tenure, $S_0$~--~spot price, $C$~--~Courant number, 
       $E$~--~error measure, $f(S_0,0)$~--~option price.
  }
  \begin{center}
  \begin{tabular}{lr|cccr|rr}
        &               & C$\approx$0.02 & C$\approx$0.01 & \multicolumn{2}{c|}{C$\approx$0.005}  && \\
$T$     &     $S_0$     &  $\log_2(E)$  &   $\log_2(E)$ &   $\log_2(E)$ &    $f(S_0,0)$ &   BS93        & European     \\\hline\hline      
0.25	&	80	&	-8.3	&	-10.1	&	-11.3	&	19.997	&	20.000	&	18.089 \\
	&	90	&	-8.1	&	-10.0	&	-11.3	&	10.036	&	10.011	&	 9.045 \\
	&	100	&	-8.0	&	-10.0	&	-11.3	&	 3.229	&	 3.162	&	 3.037 \\
	&	110	&	-8.3	&	-10.0	&	-11.3	&	 0.668	&	 0.649	&	 0.640 \\
	&	120	&	-8.2	&	-10.2	&	-11.3	&	 0.090	&	 0.087	&	 0.086 \\
\hline
0.50	&	80	&	-8.7	&	-10.2	&	-11.4	&	20.000	&	20.000	&	16.648 \\
	&	90	&	-8.5	&	-10.2	&	-11.4	&	10.290	&	10.240	&	 8.834 \\
	&	100	&	-8.5	&	-10.2	&	-11.3	&	 4.193	&	 4.109	&	 3.785 \\
	&	110	&	-8.7	&	-10.2	&	-11.4	&	 1.413	&	 1.372	&	 1.312 \\
	&	120	&	-8.6	&	-10.3	&	-11.4	&	 0.399	&	 0.385	&	 0.376 \\
\hline
3.00	&	80	&	-9.4	&	-11.4	&	-13.1	&	19.998	&	20.000	&	10.253 \\
	&	90	&	-9.3	&	-11.4	&	-13.0	&	11.697	&	11.668	&	 6.783 \\
	&	100	&	-9.3	&	-11.4	&	-13.0	&	 6.932	&	 6.896	&	 4.406 \\
	&	110	&	-9.5	&	-11.4	&	-13.1	&	 4.155	&	 4.118	&	 2.826 \\
	&	120	&	-9.3	&	-11.4	&	-13.1	&	 2.510	&	 2.478	&	 1.797 
  \end{tabular}
  \end{center}
\end{table}

\section*{Acknowledgments}

We thank Piotr Smolarkiewicz, Maciej Waruszewski, Maciej Rys and Christopher Wells for feedback on the manuscript and program code.

\section*{References}
\bibliography{paper}

\end{document}